\documentclass[%
 reprint,
 amsmath,amssymb,
 aps,
]{revtex4-2}

\usepackage{graphicx}
\usepackage{dcolumn}
\usepackage{bm}
\usepackage{colortbl}
\usepackage{xcolor}
\usepackage{amsmath,amsfonts,amsthm,amssymb} 

\begin{document}

\preprint{APS/123-QED}

\title{Dynamic Mechanisms of Catastophic Collapse:\\A New Perspective on Earthquake Physics}

\author{Klaus Regenauer-Lieb}
 \email{Klaus@curtin.edu.au}
\affiliation{%
 WASM: Minerals, Energy and Chemical Engineering, Curtin University, Perth, WA, Australia
}%


\author{Manman Hu}%
 \email{mmhu@hku.hk}
\affiliation{
 Department of Civil Engineering, The University of Hong Kong, Hong Kong
}%

\date{\today}

\begin{abstract}
The collapse of man-made and natural structures is a complex phenomenon that has been studied for centuries. We propose a new approach to understanding catastrophic instabilities, based on the idea that they do not occur at the critical point, but rather develop out of the subcritical regime as short-lived extreme events. We use an extension of Onsager's reciprocal theorem to study the subcritical regime, and we show that excitable systems in this regime are attracted to a nonlocal equilibrium that defines the maximum entropy production of at least two interacting phases. In most cases, these feedback systems are arrested by dissipative processes at larger scale, but in rare cases they can form tensor networks of instabilities that ripple from the small scale to the largest scale, forming extreme events.
\end{abstract}

\maketitle

\section{Introduction}
Earthquakes are rare events thought to involve multiple microscopic Thermo-HydroMechanical-Chemical (\textit{THMC}) processes such as interaction of fluids \citep{Miller2013}, chemical reactions or mineral breakdown \citep{Jamtveit2019} and thermally-activated friction \citep{Rice2006} driven by plate tectonic forces. The multi-scale nature of these processes makes them difficult to study. Understanding the interaction of processes at all scales is therefore still a challenge and physics-based forecasting remains a difficult proposition.   
In this work we aim at understanding the physics of instabilities in multiphase materials with internal structures that can cascade through all scales. Without restriction, we discuss geomaterials which are often conceptualized as two-phase materials (sustained by a porous skeleton) when undergoing complex multiphysics dynamic processes spanning across scales under certain natural or engineered circumstances. These scales are introduced through, for instance, system size, intrinsic discontinuities, microstructural features and the characteristic length-scales of the internal coupled processes occurring within the material during the evolution. A systematic approach underpinning the coupling of the coarse-grained dynamic process with explicit micro-scale feedbacks is desired. The development of an understanding of the complex behaviour of these systems has traditionally been the area of computational mechanics where, e.g., computational multi-scale homogenisation methods \citep{Geers2010} or simulation platforms for multiphysics phase field simulations have been developed \citep{Tonks2012}. A common drawback of these models is that update of the process information in the micro-units into the coarse-grained dynamics is not done at micro-process time, due to the intrinsic architecture of the embedding. 

Here we seek to develop an understanding from a fresh perspective based on a recently developed reaction-cross-diffusion formulation \citep{Hu2020JMPS,Geoproc2020b,Geoproc2020a,Hu2022GJI,RegenHu2023b}. This allows direct interpretation of the dynamic meso-scale processes through a thermodynamic Ansatz - in a way similar to phase field models \citep{Tonks2012} - but providing a physical meaning to the assumed internal length scale characterising the phase field. In the cross-diffusion formalism this length scale is described by meso-scale interactions resulting from the interplay between multiple processes among at least two phases. Another significant advantage of the cross-diffusion formalism is that it provides a bridge between micro-mechanical interactions and the large-scale behaviour through a semi-analytical approach enabled by its neat mathematical form.

\section{The non-local Onsager's reciprocal theorem and parity-time processes}
In an earlier contribution we have investigated multiphysics feedback processes for which the Onsager reciprocal theory provides a phenomenological approach of coupling several dissipative processes.  Onsager conceived a linear relationship between the generalised flux $J_i$ and the gradient of a generalised thermodynamic force $X_j$. The dot product of the transpose flux matrix $J^T_i$ and the force vector $X_j$ defines the local irreversible entropy production. The thermodynamic force is the thermodynamic potential difference e.g., Thermal, Hydro, Mechanical, or Chemical (\textit{THMC}) resulting in generalised thermodynamic flux described through e.g., Fourier (\textit{T})-, Darcy (\textit{H})-, Fick (\textit{C})- and solid matrix pressure (\textit{M})- diffusion laws where the diffusion coefficients ${L_{ij}}$ are denoted by the coupled processes.

\begin{equation}
J_i= -
\begin{bmatrix}\textcolor{red} {L_{TT}}&L_{TH}&L_{TM}&L_{TC}\\L_{HT}& \textcolor{red} {L_{HH}}&L_{HM}&L_{HC}\\L_{MT}&L_{MH}& \textcolor{red} {L_{MM}}&L_{MC}\\L_{CT}&L_{CH}&L_{CM}& \textcolor{red} {L_{CC}}\end{bmatrix}\  
X_j.
\label{eq:THMC}
\end{equation}

The time-symmetric state of the irreversible entropy production is described by Onsager-Casimir thermodynamic reciprocal probability statement \citep{Onsager31,Casimir} of the relaxed system (non-equilibrium steady state):
\begin{equation}
  L_{ij} = \pm {L_{ji}, (\mathrm{for}~ i \neq j)}.
\label{eq:Onsager Theorem}
\end{equation}

Onsager's reciprocal theorem \citep{Onsager31} has been used in a variety of fields such as chemistry, solid mechanics, biology and geophysical fluid dynamics. The principle delivers phenomenological constitutive laws  describing quasi-steady states of the dissipative process. For the diffusion in binary mixtures or electro-chemical systems (requiring the addition of Ohm's law for electrical diffusion (\textit{E}) extending the matrix in Eq. \ref{eq:THMC}) the linear force-flux assumption yield phenomenological diffusion coefficients which describe the dissipative structures of the system at its Minimum of Entropy Production \citep{Prigogine68}. 

The force-flux relationship in solid-like materials is, however, often nonlinear and in addition it is necessary to use Ziegler's orthogonality principle to constrain the constitutive relationship between forces and fluxes \citep{Ziegler77} defined by the Maximum Entropy Production of the product of forces and fluxes, formulating the basics  of so-called ``Thermomechanics". In this sense Prigogine's principle is a subclass of the Maximum Entropy Production principle where the vector of the Lagrange multiplier has a minimum \citep{Dewar2005} and the minimum coincides with the maximum entropy production. Ziegler's orthogonality principle is in turn a special case of a Hermitian Hamiltonian system.  Hermiticity requires that the system eigenvalues must be real; its eigenstates are orthogonal with each other and the set of all its eigenfunctions is complete. Ziegler's Thermomechanics theory is restricted to thermostatic conditions. We have recently presented an attempt to relax the thermostatic assumption by defining a thermal reference state in local equilibrium \citep{RegenHu2023a}. The extension highlights the role of the stored energy for identifying the dissipative stress as the appropriate Onsager reciprocal thermodynamic force. This consideration turns the common formulations of an asymmetric constitutive operator (non-associated plasticity) back into the classical normality condition framework \citep{RegenHu2023a}. This exercise is particularly useful for analysing the dynamic inception of large-scale earthquakes. The Hermitian assumption leads to a dissipative steady state defined by the diagonalisation of the diffusion matrix where all cross-diffusion coefficients are zero in the canonical form \citep{Meixner75}. This implies that a linearly coupled \textit{THMC(E)}  system can simply be written and solved as a set of independent equations if a local equilibrium exists and the system can be described by a Hermitian Onsager Matrix. When considering, however, the micromechanics of the dissipation process it becomes apparent that the above discussed nonlinear thermodynamic force - flux relationship incorporate a mirror reflection of mass movement, similar to the parity operator in quantum mechanics, as discussed below. They hence involve non-Hermitian matrices disguised as Hermitian matrices by using an adiabatic elimination of the cross-coupling coefficients. A closer inspection shows that in some regions of the parameter space adiabatic elimination is possible and the eigenvalues are real, however, in some other regions cross-couplings are important and eigenvalues are complex conjugate.

In order to analyse the cross-couplings from a micro-mechanical perspective we performed coarse-graining of the Langevin reaction-diffusion equation between two tightly coupled microprocesses and found it necessary to relax the local equilibrium assumption of Onsager's reciprocal theorem \citep{RegenHu2023b}. Onsager's phenomenological concept is restricted to the ideal assumption of complete isolation of the local system from its surrounding environment. However, in many \textit{THMC} systems the feedback between multiphysics process inevitably requires exchange of energy with its environment. A local microreversible system thus can at best be described as a non-local microreversible system. An example of such a non-local system might be fluid flow in a porous matrix in interaction with the compaction of the solid matrix. Using the concept of dynamic renormalisation of active zones \citep{Kardar86} we showed that there exists a diffusion regularised optimal solid-fluid transport phenomenon described by a skew symmetric Onsager constitutive operator. The skew symmetry captures the opposite directions of mass transfer between two coupled feedback processes (e.g. compaction of the solid matrix driving fluids in the opposite direction). If both processes reach a quasi-steady state over a common non-local diffusive length scale the process itself becomes independent of time. Associated with this time-symmetry there is a mirrored space symmetry, i.e. a parity operation in space where the spatial coordinates of the two interacting species have a flip in sign of the spatial coordinates. 

This is also known as a parity-time \textit{P} \textit{T} symmetry \citep{Bender98}. \textit{P} \textit{T} systems have the interesting property that there is a point in the parameter space where the real valued eigenvalues and eigenvectors of the Hamiltonian coalesce and they become complex conjugate at which point the Hermitian system becomes a non-Hermitian system. This branching from a Hermitian operator to a non-Hermitian one occurs at the ``exceptional point" \citep{Kato84} where the dynamics is not bounded by an optimization principle. The physics of the exceptional point in \textit{P} \textit{T} systems may be seen as a solution to the problem of macroscopic irreversibility stemming from microscopic reversibility also known as Loschmidt's paradox \citep{Lucia2016} as the system  can feature two different realities: (local reversible) Hermitian or (local irreversible) non-Hermetian realities depending on the value of a state variable. In solid mechanics the nucleation of complex conjugate eigenvalues has been attributed to flutter instabilities preceding macroscopic failure \citep{Loret91}. Recently, numerical experiments based on Molecular Dynamics Simulation (LAMMPS) with frictional materials revealed the exceptional point phenomenon in a solid mechanical system \citep{Chattoraj2019}. The frictional material has real valued eigenvalues at low  strain and behaves like a Hermetian system. However, at a critical strain the frictional grains exhibit an exchange of energy with their environment and the system turns into a non-Hermitian system, i.e., the nucleation of complex eigenvalues is observed. Since the solution can be written as $\exp \left(-i \lambda_n t\right)=\exp \left[-i \operatorname{Re}\left(\lambda_n\right) t\right] \exp \left[\operatorname{Im}\left(\lambda_n\right) t\right]$ the granular medium shows an oscillatory motion with an exponential growth wavelength $\lambda_n$ for any deviation from a state of mechanical equilibrium, and the second pair is an exponentially decaying wavelength. 

\section{The exceptional point in \textit{THMC} systems}
In the context of the above example the conditions for the exceptional points are spectral degeneracy and the breakdown of the diagonalisability of the Onsager operator. The possibility of the nucleation of an oscillatory instability in frictional materials at the exceptional point was proposed to be relevant to the physics of earthquakes \citep{Chattoraj2019}. The authors argue that the self-amplification of small perturbations at the exceptional point might be the micromechanical reason for the phenomenon of dynamic weakening through acoustic fluidisation, observed in geophysical experiments \citep{Giacco2015}. However, they stipulate that the study of the behaviour of frictional disks does not include the rich multiphysis of the Earth and its faults. In this contribution we put forward a proposition that can overcome this limitation. 

We suggest that it is possible to couple the multiphysics \textit{THMC} matrix through reduction to multiple degeneracy of exceptional points where the complex eigenvalues of paired microprocesses coalesce. The concatenation of bipartite couplings is an essential feature of the physics of exceptional points which are connected  through $N(N-1)$ square-root exceptional branch points in the complex plane \citep{Heiss91}. A minimum of $N=2$ processes is required with complex eigenvalues $E_m$ and $E_n$ where each of these can be connected to the next nearest singularity through further conjugate pairs of $N$ eigenvalues. 

The far from equilibrium dynamic state for nucleation of earthquake instabilities is exemplified by the \textit{THMC} coupled processes  seen as a multi-scale and multiphysics array defined by the contraction of multiple coupled processes forming a tensor network. This Riemann sheet structure of a bilinear general vector space describes the reference frame of the multiple interacting microprocesses with a symplectic basis. A system scale extreme event is triggered for the multiscale \textit{THMC} symplectic reference case where the excitation phenomena can synchronise across scale.

\begin{figure}
    \centering
    \includegraphics[width=.5\textwidth]{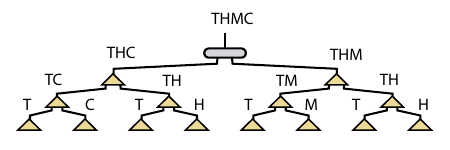}
\caption{Illustration of a multiscale hierarchical Tucker tensor network \citep{Stoudenmire2018} also known as a tree tensor network proposed for coupling of THMC reaction-diffusion transport processes. The coupled nodes represent tensors and the lines represent the binary connection forming the higher level tensor nodes. Each of the legs connects the individual branches of a Riemann sheet structure \citep{Heiss91}.}
    \label{fig:tensor network}
\end{figure}

\section{Exceptional point phenomenon as an earthquake trigger in subduction zones}
In a recent contribution we have presented the criterion for nucleation of non-Hermitian complex eigenmodes for a candidate earthquake trigger in subduction zones, based on a micro-mechanism capable of pumping fluids into the fault zone \citep{RegenHu2023b}. Coarse graining of the micromechanism resulted in an Onsager matrix with non-local cross-couplings. The cross-coupling coefficients were found to be functions of the dynamic interactions between the \textit{THMC} processes. The exact way of how the different microprocesses may interact in a multiscale scenario was not discussed.

Here, we propose to extend the theory by a multiscale tensor network \citep{Stoudenmire2018} in order to allow hierarchical connections of individual Riemann sheet structures of the multiphysics processes. The lowest level of the tensor network is defined by the individual thermodynamic  $T,H,M$ and $C$ reaction-diffusion equations. Each problem may have its own hierarchy of processes supported by variant tensor network structures, including disentangler operations between branches at each scale \citep{Stoudenmire2018}. This depends on the specific configuration, the proximity of their complex valued energy eigenstates and the growth or decay nature of their eigenstates at the exceptional point. For  the earthquake physics it may also be useful to include electrical processes as electro-chemical reactions may be important. Here, we assume the feedback of chemical (serpentinite breakdown) and a mechanical reaction (Griffith microcracking) as the resulting largest scale feedback process of the hierarchical tree. 

Choosing Fourier's law as the basis we start by coupling $HMC$- processes with a $T$-reaction diffusion equation, thus defining the dissipative ground state of the multiphysics coupled problem. This implies that other possible direct couplings like $CM$ or $HM$ request participation of temperature as a degree of freedom and are thereby implicit in the second level. Therefore, the first level of paired couplings leads to reaction-cross-diffusion equations represented by $2 \times 2$ skew symmetric tensors $TC,TH,TM$ which form the lowest level of binary non-local couplings. The second level nodes labeled $THC$ and $THM$ result in the $3 \times 3$ matrices in Eqs. \ref{eq:symplecticMatrixa} and \ref{eq:symplecticMatrixb}, which define the coupled microprocesses that may lead, for a critical state (here, a critical strain)  to an exceptional point phenomenon for nucleation of complex eigenstates of the $THMC$ problem describing the onset of an oscillatory, excited material state (here the  dilatancy pumping mechanism in subduction zones discussed later). Catastrophic collapse is here understood as an extreme event where the branch points of the lower level couplings may also have been reached, so that oscillatory instabilities support the larger scale as shown in Fig. \ref{fig:tensor network}. The \textit{THMC} coupled system is described at the highest level as a \textit{THC} and \textit{THM} coupled system with coupling between following skew-orthogonal Onsager matrices having cross-coupled nonlocal bipartite processes  $L_{TH} = - L_{HT}$ and $L_{MH} = - L_{MT}$ thus using identities of the skew coefficients the $3 \times 3$ matrices are: 

\begin{equation}
L_{ij}^{THC}=\left(\begin{matrix}0&L_{TH}&0\\-L_{TH}&0&L_{TC}\\0&-L_{TC}&0\\\end{matrix}\right),
\label{eq:symplecticMatrixa}
\end{equation}

\begin{equation}
L_{ij}^{THM}=\left(\begin{matrix}0&L_{TM}&0\\-L_{MT}&0&L_{TH}\\0&-L_{TH}&0\\\end{matrix}\right).
\label{eq:symplecticMatrixb}
\end{equation}

We can simplify further by restricting our interest to THMC processes that are precursors and may lead to the earthquake instability. For this we only consider slow slip of low magnitude and do not consider thermal-mechanical coupling that contributes to significant heating on the fault plane during rupture \citep{Braeck}. Before rupture we may assume that Onsager's  local thermal equilibrium assumption holds for the thermal state,  vindicating a Thermomechanics approach using the temperature degree of freedom as a reference frame \citep{RegenHu2023a}. We end up with a hydromechanically (HM) coupled matrix where the cross-diffusion processes involving chemical reactions can be coarse-grained into a fluid reaction term.  

This dynamic system transforms for specific dynamic Onsager coefficients into the nonlinear-Schr\"odinger equations \citep{Geoproc2020b}. These conditions are, for instance, met when fluids are captured in solid phases and are heated leading to a propensity to escape away from the heated dense solid matrix. Tight binary coupling between solid and fluid pressure $p_M$ and $p_H$ is required to enable dewatering. This is a plausible situation in a subduction environment where surface fluids are absorbed into the downgoing slab and form the solid serpentinite group minerals of a dense (no-porosity) matrix. If the slab enters the thermal regime where the mineral becomes metastable the fluid can only be released through a mechanism of tight coupling between mechanical dilatancy and fluid flow. This mechanism was first observed by Osborne Reynolds and is called dilatancy pumping \citep{Reynolds1885,Frank66}. Assuming a hydro-elastic process (Griffith microcracking opening the void space for fluids to escape) this implies that the self-diffusion coefficients   $\bar L_{ii} = 0$ of the matrix are not contributing to the process. The matrix is impermeable ($L_{HH}=0$, no Darcy diffusivity) and the viscous deformation of the matrix is not important ($L_{MM}=0$), as we are considering  a perfectly elastic micromechanism for dilatancy. A special situation arises, when  the cross-diffusion coefficients are maximised and set to opposite sign $ L_\textup{HM}^{max}= - L_\textup{MH}^{max}$ which is what is required to draw the maximum available power out of the exothermic system of serpentinite breakdown reactions at the exceptional point. This results in following equations for the evolution of solid and fluid pressures:

\begin{subequations}
\label{eq:Schroedinger2}
\begin{align}
 \frac{\partial{p_\textup{H}({\textup{\textbf{r}}, t)}}}{\partial{t}} = { L_\textup{HM}^{max} \nabla^2} p_\textup{M}(\textup{\textbf{r}}, t), \\[0.2in]
 \frac{\partial{p_\textup{M}(\textup{\textbf{r}}, t)}}{\partial{t}} = { L_\textup{MH}^{max} \nabla^2} p_\textup{H}(\textup{\textbf{r}}, t), \\[0.2in]
 \psi (\textup{\textbf{r}},t)\ = \sqrt{L_\textup{HM}^{max}}p_\textup{H}(\textup{\textbf{r}},t) - i \sqrt{L_\textup{MH}^{max}}p_\textup{M}(\textup{\textbf{r}},t).
\end{align}
\end{subequations}

where $\textup{\textbf{r}}$ denotes the position vector for the wave function $\psi (\textup{\textbf{r}},t)$ of solid and fluid pressures. As the self-diffusion coefficients are neglected the nonlinear-Schr\"odinger equations becomes an extreme form of the cross-diffusion equation. This phenomenon is well known in photonics, where it can explain optical rogue waves \citep{Akhmediev2009}, however, its analytical self-focusing 1-D solution in the $x$-direction was first given by Ref. \cite{Peregrine} who applied it to weakly nonlinear water waves. The soliton solution is:
 \begin{equation}
 \label{Peregrine}
\psi(x, t)=\left[1-\frac{4(1+2 i t)}{1+4 xi^{2}+4 t^{2}}\right] e^{i t}.
\end{equation}

\begin{figure}[htb]
	\centering
 	\includegraphics[width=.45 \textwidth]{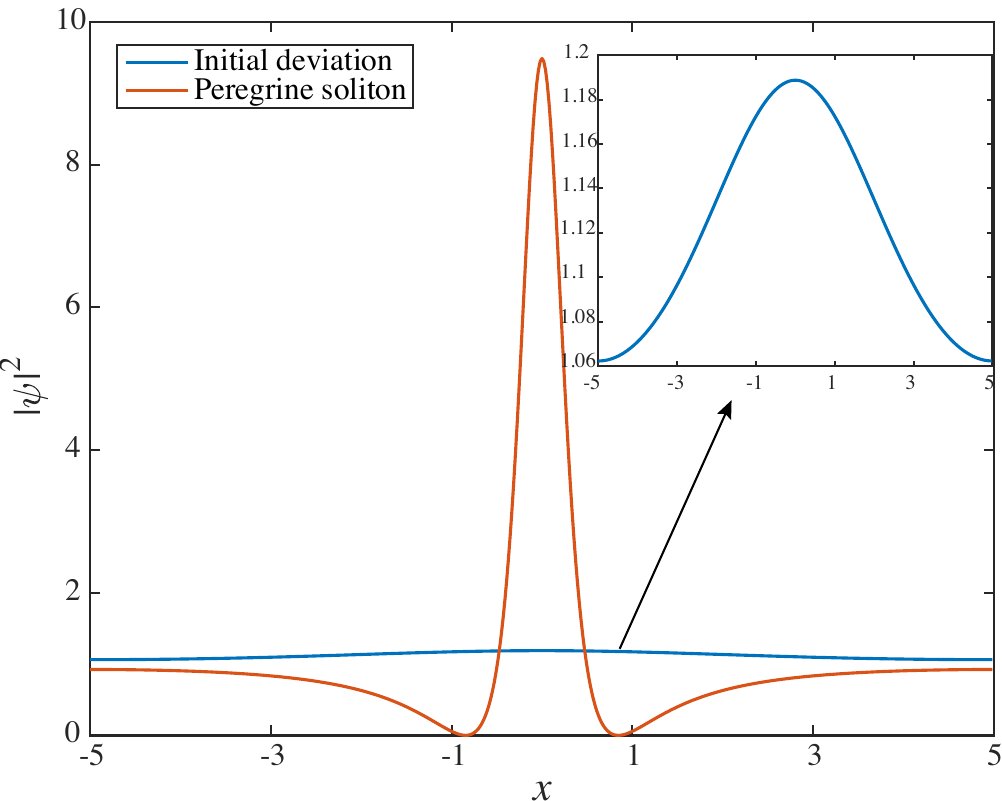}

 \caption{ Applying a small perturbation (e.g., the largest Earth  tidal signal M2)  onto a maximally cross-coupled metastable serpentinite system defined by Eq. \ref{eq:Schroedinger2}.  The resulting Peregrine soliton amplifies the power of the wave through extreme self-focusing of wave energy in space and time. The Peregrine soliton solution \citep{Peregrine} has provided a robust explanation for rogue wave observations in various materials. In the context of the dilatancy pumping mechanism the Peregrine soliton may explain the self-focusing in space and time of fluids observed through fluid induced tremor events that are related to tidal modulation and triggering of low-frequency earthquakes in northern Cascadia \citep{Royer2015}. }
\label{fig:PeregrineSoliton}	
\end{figure}

The self-focusing wave appears from low background oscillations over a very long wavelength. The rogue-wave phenomenon develops out of these localised pulses converting them into a plane wave soliton with spatial and temporal compression of wave energy by sampling energy from the far field. An illustration of such a self-focusing wave of serpentinite breakdown for a low amplitude long wavelength perturbation is shown in Fig. \ref{fig:PeregrineSoliton}. 

\section{Discussion and Conclusion}
This paper presents, to the best of our knowledge, the first attempt of explaining the earthquake mechanisms using the physics of non-local micromechanical coupling. There is therefore very limited experimental evidence for the proposed mechanism, the contribution by Ref. \cite{Chattoraj2019} is the first experimental proof. We have only discussed the highest level of feedbacks and presented assumptions for neglecting other potential exceptional point phenomena (assuming the standard local equilibrium assumption for coarse graining) such that we end up with a hydromechanical system describing the dilatancy pumping mechanism which can lead to fluid injections that are recorded as non-volcanic fluid injection (tremor) events. It is important to note that tremor events are mostly accompanied by aseismic fast episodic events and need not lead to a major earthquake \citep{Obara2002}. These slow slip and tremor events can provide a valuable data base for calibrating the models. For deciphering the multiphysics of subduction earthquakes we start by inverting the dynamic evolution of the cross-coupling coefficients allowing conditions for maximum energy release (work output) from exothermic serpentinite breakdown reactions. These conditions lead to the potential nucleation of a slab-wide synchronisation of dehydration events with the capacity to pump fluids during a short, sharp pulse into a specific location (the crest of the Peregrine wave) thus lubricating the entire downgoing oceanic slab/overriding plate interface. 

The cross-diffusion parameters can be inverted from fitting time series observations of aseismic episodic tremor and slip observations \citep{Hu2022GJI,Hu2023a} using GPS and seismological records. However, these measurements are only useful to identify the coefficient of the HM coupled matrix and the associated reactions. Because they monitor only the highest level of coupling, they have a slow-response and may have limitations for the prediction of large events. These geophysical time series record coupling of two-levels (HM) whose energy eigenstate is mostly appearing as a supercritical Hopf bifurcation with a period of 13.4 months for the Cascadia subduction case \citep{Hu2022GJI, Hu2023a}. 

For a more refined time resolution it is important to look into the energy eigenstates of the entire THMC tensor network in Fig. \ref{fig:tensor network}. An extreme event is only possible if exceptional points with positive (exponential growth) are synchronised and complex conjugate pairs that are close to each other, but can cancel each other, and disentangle the branches in Fig. \ref{fig:tensor network}, are avoided. Of particular interest for the earthquake mechanism is thus the connection of the multiscale energy levels and their potential of releasing work and triggering the global release of elastic energy stored in the plate interface on the future fault plane. Such positive multilevel coupling is known to significantly increase the amplitude \citep{Sun2021} of the Peregrine soliton wave shown in Fig. \ref{fig:PeregrineSoliton}. In order to turn the theoretical prediction into a useful forecasting tool we suggest to step away from the standard theory of self-organised critical point phenomenon that focuses on the average behaviour of the chaotic system and refer to large deviation theory \citep{Dematteis2019} which focuses on extreme events that can develop out of the subcritical regime. 

The above described mechanism of a subduction earthquake triggering event may be understood as such an extreme event of Boltzmann's theory of multinomial probabilities, including its implications for the likelihood of large deviations. It offers a fundamentally different statistical perspective to the earthquake phenomenon. The current paradigm of earthquakes as a self-organised critical point phenomenon is supported by the Gutenberg-Richter law, Omori law and the fractal nature of fault lines. The standard theory of average behaviours implies, however, that earthquakes are not predictable \citep{Bak89}. 

The proposed lens of thought offers a promising expansion of the standard theory, which is based on the simpler phenomenological angle, by considering the microphysics of the earthquake processes. Coarse-graining the additional microphysics into Onsager coefficients enables the development of a more powerful tool for analyzing general critical phenomena through self-organization. Boltzmann's theory of multinomial proabilities, including its implications for the likelihood of large deviations, offers the additional information needed for understanding how the path to criticality is achieved, inheriting information from the microstructural evolution in the subcritical thermally stable regime. Accordingly, we propose that large deviation theory \citep{Ellis99} may be used in the future for earthquake forecasting. Large deviation theory predicts exponentially fast convergence to zero probabilities of large deviations, which implies minimising the rate function of probabilities for the relative entropy also known as Sanov's theorem. It reveals that spontaneous noise induced stochastic events are possible where the Onsager cross-diffusion coefficients can either stabilise the system into supercritical Hopf bifurcations - i.e., the Episodic Tremor and Slip events as in \citep{Hu2022GJI}, or - for extremely short-lived extreme events - become the driving mechanism of the instability from the subcritical regime. In order to take the theory presented here to the next step it could be beneficial to develop an optical system of new geophysical sensors (e.g. fibre optics) capable of detecting the precursor signals to examine the hypothesis of large events nucleating out of the subcritical regime \citep{Nandan2021}.

\section{Acknowledgments} 
We acknowledge the support of the Research Grant Council of Hong Kong (ECS 27203720 and GRF 17206521) and the Australian Research Council (ARC DP170104550, DP170104557, DP200102517, LP170100233). We would also like to thank Qinpei Sun for calculations shown in Fig. \ref{fig:PeregrineSoliton}.
\newpage
\bibliography{Uncertainty}
\end{document}